\documentstyle[12pt]{article}
\normalsize

\newcounter{sxn}

\newcounter{axn}

\def\br{\begin{thebibliography}{abc}}
\def\er{\end{thebibliography}}

\date{}

\begin{document}
\bibliographystyle{unsrt}
\footskip 1.0cm
\thispagestyle{empty}

\setcounter{page}{0}
\begin{flushright}
UAHEP 9313\\
December 1993\\
\end{flushright}
\begin{center}{\LARGE Deformation Quantization \\
         of the Isotropic Rotator\\}
\vspace*{10mm}
\vspace*{6mm}
{\large   A. Stern and I. Yakushin\\}
\newcommand{\bc}{\begin{center}}
\newcommand{\ec}{\end{center}}
\vspace*{10mm}
{\it Department of Physics, University of Alabama, \\
 Tuscaloosa, Al 35487, USA.}\ec

\vspace*{5mm}

\normalsize
\centerline{\bf ABSTRACT}
We perform a deformation quantization of the classical isotropic
rigid rotator.  The resulting quantum system is not invariant under
the usual $SU(2)\times SU(2)$ chiral symmetry, but instead
$SU_{q^{-1}}(2) \times SU_q(2)$.
\newpage
\newcommand{\be}{\begin{equation}}
\newcommand{\ee}{\end{equation}}
\baselineskip=24pt
\newcommand{\ba}{\begin{eqnarray}}
\newcommand{\ea}{\end{eqnarray}}
\newcommand{\no}{\nonumber}

The classical
isotropic rotator is known to be invariant under $SU(2)\times SU(2)$
chiral transformations.
In ref. \cite {mss} a new Hamiltonian formulation of the
isotropic rotator was found where the
left and right $SU(2)$ transformations are not canonical symmetries
but rather Poisson Lie group symmetries.[2-8]  The treatment given in
ref. \cite{mss} further
differs from the standard one because the classical Hamiltonian
can not be expressed as the square of the angular momentum $J_i$, nor
does $J_i$ satisfy an $SU(2)$ algebra.  On the other hand,
from this formulation one obtains
 the usual equations of motion for the isotropic rotator.  They state
 that an $SU(2)$ matrix-valued degree of freedom $g$ denoting the
orientation of the rigid body undergoes a uniform precession.
This can be expressed as follows:
\be
\dot g g^\dagger ={i\over 2} J_i \sigma_i \;,   \qquad \dot J_i=0\;,
\ee
where the dot dentoes a time derivative,
$\sigma_i$ are Pauli matrices and we have set the moment of inertia
equal to one.  In the usual formalism a general
chiral transformation is given by
$$ g\rightarrow w^{-1}gv\;,\quad
 J_i \sigma_i \rightarrow w^{-1}J_i\sigma_i w\;,\quad
w,v\in SU(2)\;, $$
which leaves (1) invariant.

In this article we quantize the system of ref. \cite{mss}
using the method of deformation quantization \cite{bffls}.
We show that the resulting system is invariant under
$SU_{q^{-1}}(2) \times SU_q(2)$, and this is the quantum analogue of
the classical $SU(2)\times SU(2)$ Poisson Lie group symmetry.
The quantum mechanical observables for the system
are associated with a pair of Hopf algebras\cite{TT} or equivalently a quantum
double.  We obtain dynamics on the quantum double
which reduces to (1) when $\hbar\rightarrow 0$.  Furthermore in analogy
to (1) the quantum dynamics
is such that the quantum operator corresponding to
 $g$ (now taking values in $SU_q(2)$) undergoes a ``uniform precession".

We first review the classical Hamiltonian formalism of
ref. \cite{mss}.  There it was shown that the six dimensional
phase space describing a rigid
body can be taken to be the group $D=SL(2,C)$.
This phase space which is known to be a ``classical double"[2-5,8]
can be parametrized by elements of the group $G=SU(2)$ and
its dual $G^*$.  The latter is the group of $2 \times 2$ lower
triangular matrices $\{g^*\}$,
$$ g^*=\pmatrix{m &  \cr x_+  & m^{-1} \cr}\;,$$
where $m$ is real and $x_+$ is complex.
An element $\gamma$ of $D$ can be labeled by $(g^*,g)$,
$g \in G$ and $g^* \in G^*$, using the Iwasawa decomposition
 $\gamma=g^* g$.  The coordinates $(g^*,g)$
do not globally cover $D$ as, for instance, $(1,1)$ and $(-1,-1)$
are both mapped to the identity in $D$.  Nevertheless, they serve as a
useful parametrization of a finite region of $D$.

Let $e_i$, $i=1,2,3$, denote a basis for the Lie algebra ${\cal G}$
associated with $G$, and
 $e^i$ denote a basis for the Lie algebra ${\cal G}^*$
associated with $G^*$.
$e_i$ and $e^i$ together span the Lie algebra ${\cal D}$ associated
with $D$.  The Poisson brackets for $g$ and $g^*$
were expressed in terms of classical $r$-matrices, taking values in
  ${\cal D} \otimes {\cal D}$.  These matrices denoted by $r$ and $r^*$,
  were defined according to
\be
r=\; e^i \otimes e_i\;\quad{\rm and}\quad
r^*=-\;e_i \otimes e^i \;.
\ee
 $r$ and $r^*$ satisfy
\be
r^* -r= {\rm adjoint}\;{\rm invariant}\;,
\ee
and the classical Yang-Baxter equations.\cite{TT}

In this article we will make use defining representation for $D$.
In this representation the generators $e_i$ and $e^i$
can be expressed in terms of Pauli matrices $\sigma_i$ as follows:
\be
 e_i ={1\over 2} \sigma_i\;,\quad
 e^i ={1\over 2} (i\sigma_i +\epsilon_{ij3} \sigma_j)\;.
\ee
{}From them, we obtain the following $4\times4$ matrix representation
for $r$:
\be
 r={i\over 4}\pmatrix{1 & & &  \cr &-1 & &  \cr &4 &-1& \cr
& & & 1\cr} \quad ,
\ee
with $ r^*$ being is its hermitean conjugate $r^*=r^\dagger$.
Using (4), the right hand side of (3) is simply
$-{i\over 2}\sigma_i \otimes \sigma_i \;.  $

We now give the symplectic structure on $D$.
The Poisson brackets between two $SU(2)$ group elements $g$
 can be expressed according to:
\be
\{g_1,g_2\}=[\;r\;,\;g_1 g_2\;]  \;,
\ee
where we use the usual tensor product notation with
$g_1=g\otimes 1$ and $g_2=1\otimes g$.
The Jacobi identity is satisfied due to the classical Yang-Baxter
equations.  (We set the deformation parameter
$\lambda$ of ref. \cite{mss} equal to one
for simplicity.)  Poisson brackets involving group elements
$\ell^{(-)}=g^*\in G^*$ and their conjugate inverses
$\ell^{(+)}={g^{*\dagger}}^{-1}$ are given by:
\ba
\{\ell^{(\pm)}_1,\ell^{(\pm)}_2\}
&=&-[\;r^*\;,\;\ell^{(\pm)}_1 \ell^{(\pm)}_2\;]
\\
\{\ell^{(+)}_1,\ell^{(-)}_2\}&=&-[\;r^*
\;,\;\ell^{(+)}_1 \ell^{(-)}_2\;]
\ea
 The remaining Poisson brackets are
 between elements of $G$ and elements of $G^*$:
\ba
\{\ell^{(-)}_1,g_2\}&=&-\ell^{(-)}_1\; r \;g_2\;,
\\
\{\ell^{(+)}_1,g_2\}&=&-\ell^{(+)}_1\; r^* \;g_2\;.
\ea

Concerning dynamics, the classical Hamiltonian describing
a free isotropic rigid rotator on $D$ was found to be
\be
{\cal H}={1\over {2}}  Tr \ell^{(-)} {\ell^{(+)}}^{-1}\;-\;1 \;.
\ee
It leads the following Hamilton's equations of motion
\be
\dot \ell^{(\pm)} =0 \quad {\rm and} \quad
\ee
\be
\dot g g^\dagger ={i\over 2} \biggl(
 {\ell^{(+)}}^{-1} \ell^{(-)} -{1\over 2}
Tr \bigl({\ell^{(+)}}^{-1} \ell^{(-)}\bigr) \;
 {\bf 1}_{2\times 2} \biggr)
\ee
The right hand side of (13) is a traceless hermitean matrix.  We can thus
expand it in terms of Pauli matrices, identifying the coefficients
with the classical angular momenta $J_i$ of eq. (1).  This gives:
\be J_i={1\over 2 }Tr \ell^{(-)}\sigma_i  {\ell^{(+)}}^{-1}   \;.\ee
Since $ \ell^{(-)}$ and $\ell^{(+)}$ are constants of the motion,
then so are $J_i$.  Therefore the variable $g\in SU(2)$
undergoes a uniform precession, and
we recover the equations of motion for an isotropic rigid body.

As previously stated, this Hamiltonian description differs from the
usual one because the Hamiltonian (11) is not proportional to $J_iJ_I$,
and $J_i$ does not satisfy an $SU(2)$ Poisson bracket algebra.
Also different is the nature of the
 $SU(2)\times SU(2)$ chiral symmetry which we lastly review.

Unlike in the standard formulation of the rigid
rotator, the chiral transformations do not correspond to
canonical symmetries, but rather to two Poisson Lie group symmetries.
One of the Poisson Lie group symmetries
is associated with the right action of $SU(2)$ on $G$.  Elements of
$G^*$ are unchanged under such transformations.  Thus
\be
g\rightarrow gv\;, \quad \ell^{(\pm)}
\rightarrow \ell^{(\pm)}\;,\quad v\in SU(2) \;.
\ee
The Poisson brackets (6-10) for the classical
observables $g$ and $\ell^{(\pm)}$
were shown to be invariant under (15) upon insisting
 that $v$ has the following Poisson bracket with itself
\be
\{v_1,v_2\}=[\;r^*\;,\;v_1 v_2\;] \;,
\ee
and zero Poisson bracket with $g$ and $\ell^{(\pm)}$.
Then $SU(2)$ right multiplication is a Poisson map and
(15) is a Poisson Lie group transformation.
Further, since $\ell^{(\pm)}$ are unchanged by this transformation,
the Hamiltonian (11) and hence the equations of motion (12) and (13)
 are also invariant under (15).

The other Poisson Lie group symmetry
is associated with rotations as it has a nontrivial action on
$\ell^{(\pm)}$ (as well as $g$) and therefore the angular momentum $J_i$.
It corresponds to the left action of $SU(2)$.  However unlike in the
standard formalism, the left
action is not implemented directly
on $G$, but rather the classical double $D$ (and its Hermitean
conjugate).  Under the left action of $SU(2)$, the variables
$d^{(-)}=\gamma$ and $d^{(+)}={\gamma^\dagger}^{-1}$
transform according to
\be
d^{(\pm)}\rightarrow w^{-1}d^{(\pm)}\;,\quad w\in SU(2)\;.
\ee
The Poisson brackets for $d^{(\pm)}$ can be constructed from those
of $g$ and $\ell^{(\pm)}$.  One finds
\ba
\{d^{(\pm)}_1,d^{(\pm)}_2\}&=&-d^{(\pm)}_1d^{(\pm)}_2\;r-r^*\;d^{(\pm)}_1
d^{(\pm)}_2\; \;.             \\
\{d^{(-)}_1,d^{(+)}_2\}&=&-d^{(-)}_1d^{(+)}_2\;r-r\;d^{(-)}_1
d^{(+)}_2\; \;.
\ea
These relations are invariant under (17) upon insisting
 that $w\in SU(2)$ has the following Poisson bracket with itself
\be
\{w_1,w_2\}=[\;r^*\;,\;w_1 w_2\;] \;,
\ee
and zero Poisson bracket with $d^{(\pm)}$.
Then $SU(2)$ left multiplication is a Poisson map and
(17) is a Poisson Lie group transformation.
Further, since the classical Hamiltonian can be written
\be
{\cal H}={1\over {2}} Tr\; d^{(-)}{d^{(+)}}^{-1}\;-\;1 \;,
\ee
we can use the cyclic property of trace to show that it
is unchanged under (17).

We are now ready to consider the quantization of this system.
In the spirit of deformation quantization
 \cite{bffls}, we do not identify the quantum mechanical
commutation relations with $i\hbar$ times the corresponding
classical Poisson brackets, but only demand that they agree
in the $\hbar\rightarrow 0$ limit.
Also, in the spirit of deformation quantization, we do not identify the
quantum Hamiltonian ${\bf H}$ with the classical Hamiltonian ${\cal H}$.
Rather we only require that ${\bf H}$
 reduces to ${\cal H}$ (with classical variables replaced by quantum
operators) in the limit $\hbar\rightarrow 0$.
These requirements are of course
not enough to completely determine the quantum
mechanical system.  For this purpose we shall in addition insist
that the Heisenberg equations governing the quantum dynamics are the
quantum analogs of the classical equations of motion (12) and (13),
and that for each Poisson Lie group symmetry present in the
classical theory there is a corresponding quantum symmetry.
Regarding the former, we want that the quantum operators corresponding
to $ \ell^{(\pm)}$ are constants of the motion, and the quantum operator
corresponding to $g$ undergoes a ``uniform precession".
Concerning the latter, the resulting symmetry transformations
are not associated with groups, but Hopf algebras.\cite{TT}

We begin by writing down the quantum mechanical commutation relations.
The Poisson bracket algebra (6) for $g$ is known to
be identical to the semiclassical limit of the $SU_q(2)$
Hopf algebra.  $SU_q(2)$ can be described in terms of $2\times 2$
matrices $\{T\}$ whose matrix elements are not c-
numbers, but rather satisfy the commutation relations:
\be
R T_1T_2=T_2 T_1 R
\ee
 with $T_1=T\otimes 1$, $T_2=1\otimes T$ and $R$ given by
\be
R=q^{-1/2}\pmatrix{q & & & \cr & 1 & & \cr & q-q^{-1} &1 & \cr
& & & q \cr}\;.
\ee
In addition,
 $T^{\dagger} T={\bf 1}_{2\times 2}$ and $det_q T=T_{11}T_{22}
-qT_{12}T_{21}=1$.
$R$ satisfies the quantum Yang-Baxter equation, as well as
\be
[RR^T\;,T_2T_1\;] =0
\ee
which is the quantum analogue of the condition (3).  The latter relation
can be verified using the $2\times2$ matrix representation for $T$.
Here $q=e^{\hbar / 2}$.  In the $\hbar\rightarrow 0$ limit
 $R$ tends to $ 1 - i \hbar r + O(\hbar^2)$, and consequently
(22) reduces to $[\;T_1, T_2\;] =i\hbar[\;r\;,\;T_1 T_2\;]+ O(\hbar^2) $.
  We thereby recover the algebra given in (6).

The Poisson bracket algebra for $\ell^{(\pm)}$ given in (7) and (8)
is known to be identical to the semiclassical limit of the $U_q(sl(2))$
Hopf algebra.  $U_q(sl(2))$ can be described by the set of $2\times 2$
lower triangular matrices $\{L^{(-)}\}$
and upper triangular matrices $\{L^{(+)}\}$ given by
\be
L^{(-)} =\pmatrix{q^{-H/2} & \cr -(q-q^{-1})X_+  & q^{H/2} \cr}
\quad{\rm and}\quad
L^{(+)} =\pmatrix{q^{H/2} & (q-q^{-1})X_-\cr  & q^{-H/2} \cr}\;.
\ee
The commutation relations for $H$, $X_+$ and $X_-=\bar {X_+}$ are
\be
 [  H,X_\pm ]  =\pm 2X_\pm\;,
\quad{\rm and}\quad
 [ X_+,X_- ]  ={{q^H - q^{-H}}\over{q-q^{-1}}}\;,
\ee
or equivalently,
\ba
R^{(+)} L^{(\pm)}_1L^{(\pm)}_2&=&L^{(\pm)}_2 L^{(\pm)}_1 R^{(+)}\\
R^{(+)} L^{(+)}_1L^{(-)}_2&=&L^{(-)}_2 L^{(+)}_1 R^{(+)}
\ea
with $L^{(\pm)}_1=L^{(\pm)}\otimes 1$,
$L^{(\pm)}_2=1\otimes L^{(\pm)}$ and $R^{(+)}=R^T$.
In addition to the identity (24) we have
$[RR^T\;,L_1^{(\pm)}L_2^{(\pm)}\;] =$
$[RR^T\;,L_1^{(+)}L_2^{(-)}\;] =0$.
In the $\hbar\rightarrow 0$ limit
$R^{(+)}$ tends to $1+ i\hbar r^* + O(\hbar^2)$
and consequently we recover the classical algebra (7)
and (8) from (27) and (28).

Since the Poisson brackets between $g$ and $\ell^{(\pm)}$
do not vanish, it follows that the elements $T\in
SU_q(2)$ do not commute with the elements $L^{(\pm)}\in
U_q(sl(2))$ in the quantum theory.  The quantum mechanical
commutation rules for $T$ with $L^{(\pm)}$
must correspond to the Poisson brackets (9) and (10)
in the limit $\hbar \rightarrow 0$.  A suitable set of commutation
relations consistent with this requirement is
\be
 L^{(\pm)}_1 R^{(\pm)} T_2=T_2 L^{(\pm)}_1\;,
\ee
where $R^{(-)}=R^{-1}$.
In the limit $\hbar\rightarrow 0$,
$R^{(+)}$ and $R^{(-)}$ tend to $1+ i\hbar r^* + O(\hbar^2)$ and
$ 1+i\hbar r+O(\hbar^2)$, respectively, and the Poisson brackets
(9) and (10) are recovered from (29).

The commutation relations for $L^{(\pm)}$ and $T$ are completely
determined by eqs. (22,27-29).  With the use of the quantum Yang-Baxter
relations it can be checked that no further conditions
on $L^{(\pm)}$ and $T$ result from
commuting $L^{(\pm)}$ and $T$ through (22,27-29).

We next determine the quantum Hamiltonian ${\bf H}$.  Since we insist
that the Heisenberg equations governing the quantum dynamics correspond
with the classical equations of motion (12) and (13),
 $ L^{(\pm)}$ should be constants of the motion,
and $T$ should undergo a ``uniform precession".
For the former we have that $L^{(\pm)}$ commutes with ${\bf H}$,
and thus ${\bf H}$ must be a function of only the Casimir ${\cal C}$ for
$U_q(sl(2))$.  The latter is known to be
\be
{\cal C}= Tr_q \;L^{(-)} S(L^{(+)})
=q^{H+1} +q^{-H-1} +(q-q^{-1})^{2} X_- X_+ \;,
\ee
where $S(A)$ denotes the antipode of $A$.  For both Hopf algebras
$SU_q(2)$ and $U_q(sl(2))$ the antipode
 is known to behave like a matrix inverse, ie. $S(A)A=AS(A)=
 {\bf 1}_{2\times 2} $.
$S(L^{(\pm)})$ are given in terms of $2\times2$ matrices according to
\be
S(L^{(-)}) =\pmatrix{q^{H/2} & \cr (1-q^{-2})X_+  & q^{-H/2} \cr}
\quad{\rm and}\quad
S(L^{(+)}) =\pmatrix{q^{-H/2} & -(q^2-1)X_-\cr  & q^{H/2} \cr}\;.
\ee
$Tr_q$ in eq. (30) denotes a ``quantum" trace.\cite{tr}
$Tr_q$ of a $2\times 2$ matrix $M= [ M_{ij} ]$ is defined according to:
\be
Tr_q \;M=q M_{11}+ q^{-1}M_{22}\;.
\ee
Unlike the usual trace, $Tr_q$ does not have the general
property of invariance under cyclic permutations.  It does however
serve as an ``adjoint invariant" with respect to both
Hopf algebras $SU_q(2)$ and $U_q(sl(2))$.  By this we mean the following:
\ba
Tr_q \;L^{(\pm)} M S(L^{(\pm)})& =& Tr_q\;M\;, \qquad
 [ L^{(\pm)}_1,M_2 ]  =0 \;,       \\
Tr_q \;S(T) M T& =& Tr_q \;M\;, \qquad [ T_1,M_2 ]  =0 \;.
\ea
These relations can be explicitly verified using the $2\times2$
representations for $L^{(\pm)}$ and $T$.
{}From the requirement that the quantum Hamiltonian ${\bf H}$
 reduces to ${\cal H}$ in the limit $\hbar\rightarrow 0$, we can choose
\be
{\bf H}={1\over {2}}{\cal C}-1 \;.
\ee
(More generally we can add terms to (35) that are of order $\hbar$.
We shall not consider that possibility here.)

To compute the equation of motion for $T\in SU_q(2)$ we take its
commutator with ${\cal C}$.  Using (29), we find
\be
 [ {\cal C},T_2 ]  =Tr^1_q\; L_1^{(-)}
 \biggl(1-(R^TR)^{-1}\biggr) S(L^{(+)})_1 T_2
\;,
\ee
where the ``$1$" index in
$Tr_q^1$ indicates that the ``trace" is performed only on
the first space in the tensor product.
So from the Heisenberg equation of motion
$\dot T_2 ={i\over \hbar}[ {\bf H},T_2 ] $, we get
\be
\dot T_2 S(T)_2 ={i\over {2\hbar}}
Tr_q^1\; L_1^{(-)} \biggl(1-(R^TR)^{-1}\biggr) S(L^{(+)})_1  \;.
\ee
Some work shows that this equation can be rewritten according to
\be
-2i\hbar \; \dot T S(T) =
(1-q^{-2})  S(L^{(+)}) L^{(-)}  + (1-q){\cal C}\; {\bf 1}_{2\times 2}
\;.
\ee
Since the right hand side of the Heisenberg equation of motion
(38) is a function of $L^{(\pm)}$ only
it is a constant of the motion, just as is the right hand side of
the classical equation of motion (13).  We therefore conclude that
in analogy to $g$, $T$ undergoes a ``uniform precession".
In the $\hbar\rightarrow 0$ limit, eq. (38) reduces to
\be
-2i\;\dot T S(T) \rightarrow
  S(L^{(+)}) L^{(-)}  -{1\over 2}
Tr \bigl( L^{(-)}S( L^{(+)})\bigr) \;   {\bf 1}_{2\times 2}
\ee
which agrees with the classical equation of motion (13).

We finally show that the above system is invariant under
$SU_{q^{-1}}(2) \times SU_q(2)$.  This is the quantum analog of the
$SU(2)\times SU(2)$ Poisson Lie group symmetries of the classical theory.
One of the Poisson Lie group symmetries
is associated with the right action of $SU(2)$ on itself given in (15).
The corresponding symmetry transformation in
the quantum theory is the right action of $SU_q(2)$
on itself.  Elements of $U_q(sl(2))$
are unchanged under these transformations.  Thus
\be
T\rightarrow TV\;, \quad L^{(\pm)}\rightarrow L^{(\pm)}\;,
\quad V\in SU_q(2) \;.
\ee
The commutation relations for the quantum mechanical
observables $T$ and $L^{(\pm)}$ are invariant
under (40) if we insist that $V$ satisfies the $SU_q(2)$
commutativity relation: $R V_1V_2=V_2 V_1R \;.$
For this we also assume that $V$ commutes with $T$ and $L^{(\pm)}$.
Since $L^{(\pm)}$ are unaffected by the action of $SU_q(2)$,
the Hamiltonian and hence the Heisenberg
equation of motion are unchanged under (40).

The other Poisson Lie group symmetry
was associated with the left action of $SU(2)$ on $D$ given in (17).
In order to find the corresponding symmetry transformation in
the quantum theory we first need the analogue of the classical
observables $d^{(\pm)}$.  For this we define
\be
D^{(\pm)}=L^{(\pm)} T\;.
\ee
{}From the commutation relations for $L^{(\pm)}$ and $ T$ we obtain:
\ba
R^{(+)} D^{(\pm)}_1D^{(\pm)}_2&=&D^{(\pm)}_2 D^{(\pm)}_1 R\;,\\
R^{(-)} D^{(-)}_1D^{(+)}_2&=&D^{(+)}_2 D^{(-)}_1 R\;.
\ea
The symmetry transformation in
the quantum theory associated with (17) is
\be
D^{(\pm)}\rightarrow S(W)D^{(\pm)}\;,\quad W\in SU_q(2)\;,
\ee
where $R W_1W_2=W_2 W_1R \;, $
and we also assume that $W$ commutes with $D^{(\pm)}$.
Since $S(W)$ satisfies the $SU_{q^{-1}}(2)$ commutation relations,
we say that the transformation (44) is the left action of
$SU_{q^{-1}}(2)$.
To show that the commutation relations (42) and (43)
for $D^{(\pm)}$ are unchanged under (44) we can use
$[R^-{R^-}^T\;,S(W)_2S(W)_1\;] =0$ which follows from (24) (with $T$
replaced by $W$).
The quantum Hamiltonian and the Heisenberg equation of motion
can be written in terms of $D^{(\pm)}$ according to
\be
{\bf H}={1\over 2} Tr_q\;\Gamma \; \;-\;1
\;,\qquad \Gamma=  \;D^{(-)} S(D^{(+)})     \;,
\ee
and
\be
-2i\hbar\;\dot D^{(-)}  S(D^{(-)}) =
(1-q^{-2}) \;\Gamma  + (1-q) \;
 Tr_q \;\Gamma\; \;{\bf 1}_{2\times 2} \;,\qquad \dot \Gamma=0\;,
\ee
respectively, where we use $ S(D^{(\pm)}) = S(T) S(L^{(\pm)})$.  They
are unchanged under (44) due to the property (34) of the deformed trace.

{\bf Acknowledgements}

We were supported in part by the Department of Energy, USA under
contract number DE-FG-05-84ER-40141.  A. S. wishes to thank the
group in Naples for their hospitality where this work was initiated.
We are grateful for discussions with S. Rajeev.

\end{document}